\documentclass[prb,twocolumn,showpacs,showkeys,amssymb]{revtex4}

\usepackage{graphicx} 
\usepackage{dcolumn} 
\usepackage{bm} 

\begin{document}

\title{Selection and jump rules in electronic Raman scattering from GaAs/Al$_{x}$Ga$_{1-x}$As artificial atoms}
\author{Alain Delgado$^a$, Augusto Gonzalez$^b$, and D.J. Lockwood$^c$}
\affiliation{$^a$Centro de Aplicaciones Tecnol\'ogicas y
 Desarrollo Nuclear, Calle 30 No 502, Miramar, Ciudad Habana, C.P. 11300, Cuba\\
 $^b$Instituto de Cibern\'etica, Matem\'atica y F\'{\i}sica, Calle
 E 309, Vedado, Ciudad Habana, Cuba\\
 $^c$Institute for Microstructural Sciences, National Research Council,
 Ottawa, Canada K1A 0R6}

\begin{abstract}
A theoretical description of electronic Raman scattering from GaAs/Al$_{x}$Ga$_{1-x}$As artificial atoms under the influence of an external magnetic field is presented. Raman spectra with laser excitation energy in the interval $E_{\rm{gap}}-30 \hspace{.1cm} \rm{meV}$ to $E_{\rm{gap}}$ are computed in the polarized and depolarized geometry. The polarization ratios for the collective and single-particle excitations indicate a breakdown of the Raman polarization selection rules once the magnetic field is switched on. A Raman intensity jump rule at the band gap is predicted in our calculations. This rule can be a useful tool for identifying the physical nature (charge or spin) of the electronic excitations in quantum dots in low magnetic fields. 
\end{abstract}


\maketitle

The spectrum of electronic excitations of quantum dots is, in many cases, required
as basic information for developing relevant applications. By performing inelastic light (Raman) scattering experiments, one can examine a whole sector of elementary excitations not accessible through photoluminescence or absorption measurements. In the 1990s, Raman experiments on GaAs/GaAlAs deep-etched quantum dots reported interesting results. An overview that summarizes this work can be found in Ref. [\onlinecite{Lockwood_review_2000}]. 

Sch\"uller {\it et al.}\cite{shuller_PRB_1996, shuller_PRL_1998} obtained Raman spectra dominated by collective excitations by illuminating the samples with lasers whose energy values were well (40 meV) above the effective band-gap of the dot. Two types of collective modes were found; the charge-density excitations (CDEs) and the spin-density excitations (SDEs) and these can be distinguished by polarization selection rules. CDEs are observed when the polarizations of the incident and scattered light are parallel to each other (polarized geometry), while SDEs are detected in a configuration where the scattered light polarization is perpendicular to the incident one (depolarized geometry). These polarization selection rules are obtained in the framework of the off-resonant approximation (ORA)\cite{ORA_PHYE_2004} where the Raman transition amplitude is directly determined by the multipole strength functions of the final excited states. The authors in Ref. [\onlinecite{ORA_PHYE_2004}] investigate the limits to the general use of the ORA depending on the quantum numbers of the final excited states and the characteristic lateral confinement energy of the quantum dot. For a laser excitation energy, $h\nu_{i}$, below band-gap the off-resonant regime is only reached for monopolar and quadrupolar modes when $h\nu_{i}\le E_{\rm{gap}}-30$ meV \cite{ORA_PHYE_2004}. Strong Raman peaks, appearing in both geometries, when the energy of the incident light is quite close to the effective band-gap (extreme resonance regime), were interpreted by Lockwood {\it et al.} \cite{lockwood_PRL_1996} in terms of single-particle excitations (SPEs). More recently, Brocke {\it et al.} \cite{heitmann_PRL_2003} reported the first observations of a few-electron CDEs in InGaAs self-assembled quantum dots by means of Raman spectroscopy.

In this paper, we compute the near-resonance Raman spectrum of a medium-size disk-shaped quantum dot containing 42 electrons in an external magnetic field applied perpendicular to the plane of the dot. The method and approximations used to obtain the many-body wave functions and energies of the required states are explained in Ref. [\onlinecite{AAL_PRB_2004}]. Energies and wave-functions of the final excited states are obtained by means of the random phase approximation \cite{peter_ring_book} (RPA). The calculation of the monopole operator matrix elements between each final state and the ground state, allows the identification of the collective modes (CDEs, SDEs) and the SPEs. By computing the Raman polarization ratios we evaluate the extent to which the magnetic field breaks down the polarization selection rules for both collective and SPEs. We report the first observation of what we term the Raman
intensity jump-rule. This effect, more clearly seen for the collective modes, can be enunciated as follows: a monotonic increase of a Raman peak intensity 
when the incident laser energy is increased from below up to the effective band-gap of the dot in the polarized (depolarized) geometry indicates that the excited state associated with that peak is a charge (spin) excitation. If we look at the same state in its opposite geometry, we find quite small values of the Raman intensities at any incident energy except at the effective band-gap, in which case a sudden jump of the Raman intensity should be detected.

From theory \cite{Loudon_text}, we know that the Raman peak intensity is proportional to the transition amplitude squared, $\vert A_{fi}\vert ^{2}$, where

\begin{equation}
A_{fi}\sim \sum_{int} \frac{\langle f|\hat H^+_{e-r}|int\rangle
\langle int|\hat H^-_{e-r}|i\rangle}{h\nu_i-(E_{int}-E_i)+i\Gamma_{int}}.
\label{eq1}
\end{equation}
For simplicity, the initial, intermediate and final quantized states of the radiation field are not written explicitly. $|i\rangle$, $|int\rangle$ and $|f\rangle$ are, respectively, the initial (ground), intermediate and final (excited) electronic states of the quantum dot. $\hat H_{e-r}$ is the electron-radiation interaction Hamiltonian \cite{ORA_PHYE_2004}, and $\Gamma_{int}$ is a phenomenological damping parameter. A consistent calculation of the Raman intensity requires reliable approximations to the many-particle wave functions $|i\rangle$, $|int\rangle$, and $|f\rangle$. Since we are considering incident laser energies ($h\nu_{i}$) of the order of the effective band-gap of the quantum dot, as is the case for most experimental work, intermediate states with an additional electron in the conduction band and one hole in the valence band will be the relevant contributors to the summation in Eq. (\ref{eq1}). The matrix element $\langle int|\hat H^{-}_{e-r}|i\rangle$ represents the virtual transition from the $N_{e}$-electron ground state to an intermediate state in which an incident photon is annihilated and an electron-hole pair is created. The inverse mechanism is described by the matrix element $\langle f|\hat H^{+}_{e-r}|int\rangle$, where the electron-hole pair is annihilated and a scattered photon of energy $h\nu_{f}<h\nu_{i}$ is emitted. 

We modelled our GaAs quantum dot with a disk of thickness $L=25$ nm. The electron motion in the growth direction is confined by a hard wall potential. On the other hand, the lateral confining potential is assumed to be parabolic with a characteristic energy $\hbar \omega_{0}=12$ meV. Hole states in the valence band are found by diagonalizing the Kohn-Luttinger Hamiltonian \cite{KL_matrix_elements} in the presence of an external confinement and the $42$ electron background.

We focus on monopolar excitations. This means that the total angular momenta of the 
initial and final states are equal. Collective excitations (CDEs, SDEs) correspond to $|f\rangle$ states giving a significantly nonzero value of the matrix element:

\begin{equation}
D^0_{fi}=\left\langle f\left| \sum_{\alpha,\beta}
 d^0_{\alpha,\beta}[ \hat e^\dagger_{\alpha\uparrow}\hat e_{\beta\uparrow} \pm
 \hat e^\dagger_{\alpha\downarrow}\hat e_{\beta\downarrow} ]
 \right|i\right\rangle,
 \label{multipole_collective}
\end{equation}
where $d^0_{\alpha,\beta}$ is the matrix element of the monopole operator, defined in Ref. [\onlinecite{AAL_PRB_2004}]. The greek subindices denote the orbital part of the HF electronic states, while the spin projection is explicitly indicated by the arrow. The plus (minus) sign in Eq. (\ref{multipole_collective}) corresponds to the CDEs (SDEs).

\begin{figure}[ht]
\begin{center}
\includegraphics[width=1.0\linewidth,angle=-90]{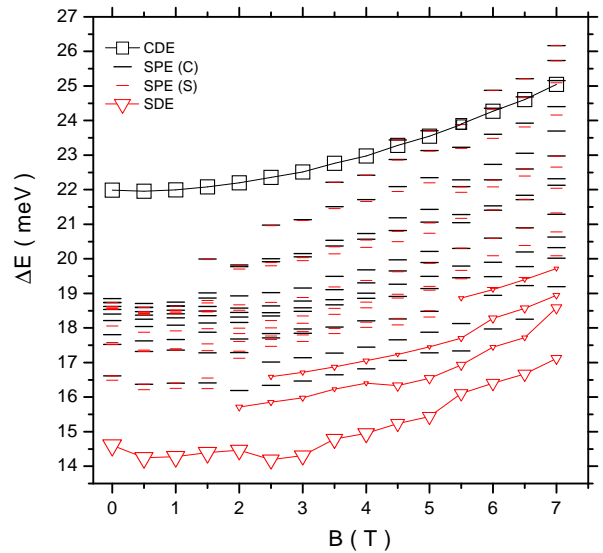}
\caption{\label{fig1} (Color online) Monopolar excitation spectrum of the dot.}
\end{center}
\end{figure}

Figure \ref{fig1} displays the monopolar excitation spectrum of the modelled dot in an external magnetic field. CDEs (open squares) and SDEs (open triangles), whose contribution to the energy-weighted sum rules (EWSRs) \cite{AAL_PRB_2004} exceeds 5 \%, are specially indicated. The contribution to the EWSR of each collective mode is shown by means of the size of the symbols. Charge (SPEs(C)) and spin (SPEs(S)) single-particle excitations are also represented with long and short horizontal bars, respectively. For the CDE mode, one notices two features: i) its energy evolves in a
smooth way as $B$ is increased, and ii) the square sizes do not change with the field. This latter fact indicates the strong collective nature of the mode, which concentrates all of the oscillator-strength. Only at $B=5.5$ T are there two collective states sharing the whole contribution to the EWSR. 
The lowest excitation energies in Fig. \ref{fig1} correspond to the SDEs. One can
see at $B=2$, $2.5$, and $5.5$ T starting points for new SDE bands. The lowest energy band  shows no significant dependence on the magnetic field up to $B=2$ T, while at higher fields the SDEs follow a quasi-linear dependence. Notice how the oscillator-strength of the SDEs, concentrated in one state up to $1.5$ T, is redistributed between more than one mode for higher values of $B$. SPEs are in between
the SDEs and the CDE. They are grouped in compact bunches of states at low values of $B$. When $B\geq 3$ T a more homogeneous distributions of these states is found. Notice that the first single-particle excited state (spin or charge) is relatively isolated from the rest for any value of the magnetic field. 

\begin{figure}[ht]
\begin{center}
\includegraphics[width=1.0\linewidth,angle=0]{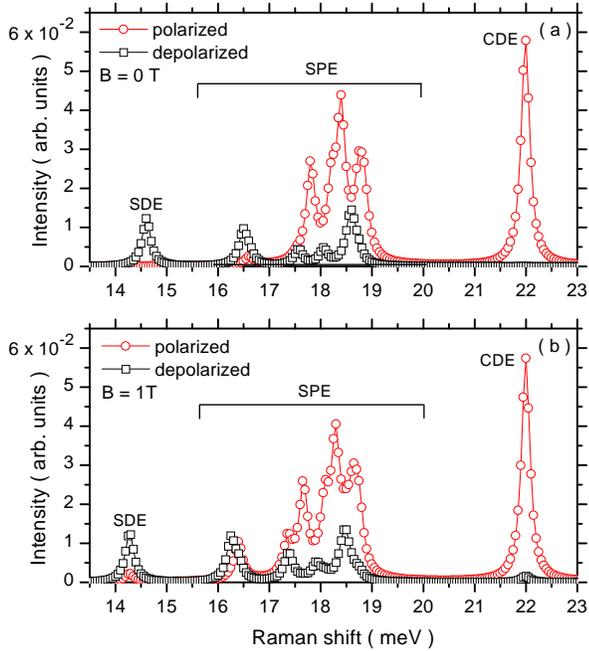}
\caption{\label{fig2} (Color online) Polarized and depolarized Raman spectra at $B=0$ and $1$ T. The incident laser energy is $h\nu_{i}=E_{\rm{gap}}-2.5$ meV.}
\end{center}
\end{figure}

The Raman spectra we present in this paper are computed in a backscattering configuration where the incident and scattering angles, defined with respect to the normal of the dot are fixed at $20^{\rm{o}}$. The incident laser energy is measured
with respect to the effective band-gap of the dot, $E_{\rm{gap}}=E^{(0)}_{int}-E_{i}$, where $E^{(0)}_{int}$ is the lowest intermediate state energy, and $E_{i}$ -- the
ground state energy. Each intermediate state entering Eq. (\ref{eq1}) satisfies the
condition $E_{int}-E^{0}_{int}\leq \hbar\omega_{\rm{LO}}$, with $\hbar\omega_{\rm{LO}}\sim 30$ meV being the threshhold excitation energy for the spontaneous emission of longitudinal optical phonons in GaAs. In this interval, the phenomenological parameter,
$\Gamma_{int}$, related to the lifetime of the $|int\rangle$ states, is fixed at $0.5$ meV. For higher excitation energies, the emission of LO phonons becomes active and the widths of those states should experience a sudden increase \cite{phonons}. This regime is not considered in the present paper.
Calculated monopolar Raman spectra in both geometries at $B=0$ and $1$ T are shown in Fig. \ref{fig2}. The energy of the incident laser is  $2.5$ meV below the band gap. The Raman peaks related to the SDE, SPEs and CDE are explicitly indicated. Figure \ref{fig2}(a) reveals an SDE peak which is only visible in the depolarized geometry. The situation for the CDE peak is completely the opposite; a non-zero Raman intensity is only observed in the polarized geometry. However, once the magnetic field is switched on this polarization selection rule is partially broken. As can be seen in Fig. \ref{fig2}(b), non-trivial values of the Raman intensities associated with the CDE and SDE are obtained in both geometries, although the polarized (depolarized) geometry remains the favorable configuration for detecting the CDE (SDE). 

The polarization ratio $r$ of a Raman peak related to a charge (collective or single-particle) excitation is defined by the fraction whose numerator is the squared Raman amplitude computed in the depolarized geometry, which is the unfavorable geometry for this mode, and the denominator is the squared amplitude in the polarized geometry. For spin excitations the fraction is inverted. At $B=0$, the polarization ratios of the CDE and SDE are $r^{B=0\rm{T}}_{\rm{CDE}}=3.6\times 10^{-4}$ and $r^{B=0\rm{T}}_{\rm{SDE}}=2.0\times 10^{-7}$. In a magnetic field of $1$ T, the ratios reach the values $r^{B=1\rm{T}}_{\rm{CDE}}=2.8\times 10^{-2}$ and $r^{B=1\rm{T}}_{\rm{SDE}}=1.7 \times 10^{-1}$. Notice the strong perturbing effects of the external magnetic field on the Raman polarization selection rules, especially for the SDE. 

\begin{figure}[ht]
\begin{center}
\includegraphics[width=1.0\linewidth,angle=0]{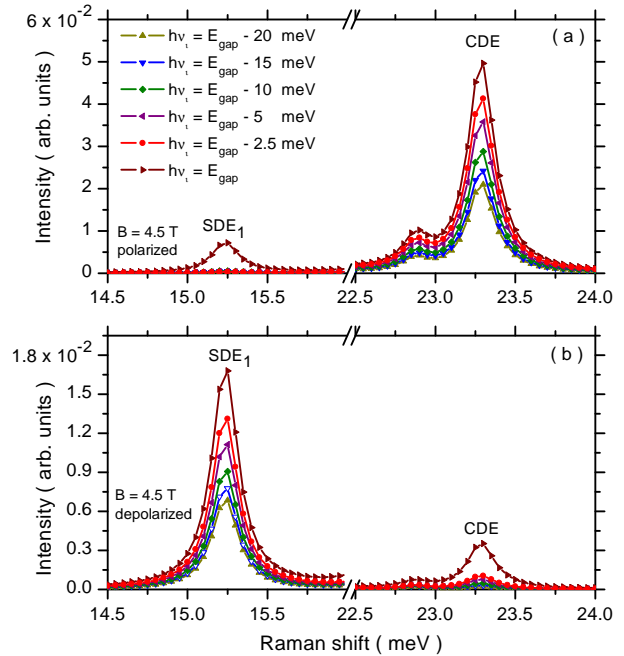}
\caption{\label{fig3} (Color online) The collective Raman peaks at $B=4.5$ T as $h\nu_{i}$ approaches the band gap.}
\end{center}
\end{figure}

SPEs with excitation energies ranging from $16$ to $19$ meV are apparent in Fig. \ref{fig2}. At $B=0$ and $1$ T, the SPEs(C) and SPEs(S) are quite close in energy and the overlapping intensities make a separate qualitative experimental analysis difficult. However, we can compute $r$ for each SPE. At $B=0$, the squared amplitudes of SPEs(C) in the depolarized geometry are, on average, three orders of magnitude smaller than in the polarized geometry, giving an average polarization ratio $\langle r^{B=0\rm{T}}_{\rm{SPE(C)}} \rangle=3.0 \times 10^{-3} $. This indicates that monopolar Raman peaks related to SPE(C) obey, by analogy with the CDEs, a polarization selection rule at $B=0$ T. At $B=1$ T, however, the Raman amplitudes exhibit 
comparable magnitudes in both geometries, giving an average polarization ratio, $\langle r^{B=1\rm{T}}_{\rm{SPE(C)}}\rangle=0.26$. For the SPEs(S), a transition from $\langle r^{B=0\rm{T}}_{\rm{SPE(S)}}\rangle=2.1\times 10^{-4}$ to $\langle r^{B=1\rm{T}}_{\rm{SPE(S)}}\rangle=1.1$ is found. 
 
\begin{figure}[ht]
\begin{center}
\includegraphics[width=1.0\linewidth,angle=0]{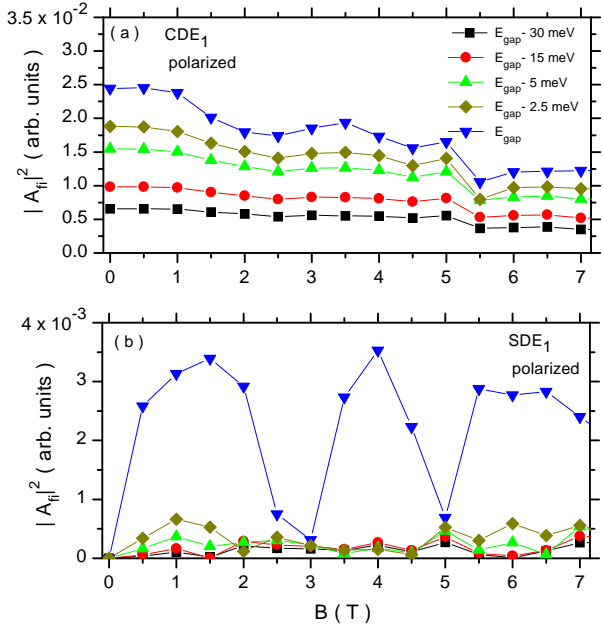}
\caption{\label{fig4} (Color online) Squared Raman amplitudes related to the (a) CDE and 
(b) $\rm{SDE_{1}}$ in the polarized geometry versus the external magnetic field.}
\end{center}
\end{figure}

The jump rule is depicted in Fig. \ref{fig3}. We focus on the behavior of the Raman peaks linked to the collective modes. 
$\rm{SDE}_{1}$ labels the lowest spin excitation. Fig. \ref{fig3}(a) illustrates how the polarized Raman intensities change when the energy of the incident laser is increased from $h\nu_{i}=E_{\rm{gap}}-30$ meV up to $E_{\rm{gap}}$. A monotonic increase of the Raman intensity related to the CDE is apparent. However, for the SDE we find quite small values of the Raman intensities except at the band-gap, where a sudden jump in the intensity is observed. Fig. \ref{fig3}(b) shows results in the depolarized geometry, the Raman peak that experiences a monotonic increase is the $\rm{SDE}_{1}$, while a jump rule is observed for the CDE. The jump rule involves two factors: the character (charge or spin) of the electronic excitation associated with a given Raman peak, and the polarization in which the Raman intensity is computed (or measured). 

It is useful to examine the validity of the jump rule at different magnetic field values. Fig. \ref{fig4} shows the variation of the Raman intensities of the $\rm{CDE}$ and $\rm{SDE_{1}}$ with the magnetic field in the polarized geometry. Results are shown for several values of the incident light energy. A monotonic dependence of the CDE Raman intensity is observed. For the $\rm{SDE_{1}}$, a jump rule picture is obtained at any field value except at $B=2.5$, $3.0$ and $5.0$ T. Notice at $B=0$ T that the conventional polarization selection rules are fulfilled independently of the incident laser energy.

\begin{figure}[ht]
\begin{center}
\includegraphics[width=1.0\linewidth,angle=0]{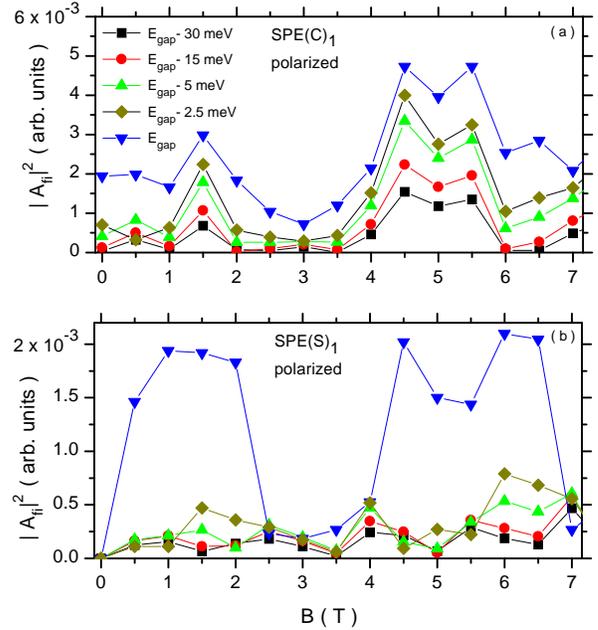}
\caption{\label{fig5} (Color online) The same as Fig. \ref{fig4}, but for the (a) $\rm{SPE(C)_{1}}$ and 
(b) $\rm{SPE(S)_{1}}$ peaks.}
\end{center}
\end{figure}

In Fig. \ref{fig5}, we present a similar analysis  for the Raman intensities associated with the $\rm{SPE(C)_{1}}$ and  $\rm{SPE(S)_{1}}$. From an experimental point of view, the Raman peaks related to these single-particle excited states, which are relatively isolated from the others, can in principle, be followed as a function of the magnetic field. Notice in Fig. \ref{fig5}(a) that the Raman intensities vary strongly when the magnetic field is changed, and the monotonic increase as a function of the laser energy. On the other hand, for the $\rm{SPE(S)_{1}}$, we find two magnetic field intervals where a jump rule picture applies. 

The Raman intensity jump rule results from the breakdown of the polarization selection rules at the band gap excitation, where the ORA is no longer valid. We have examined this interesting effect as a function of the external magnetic field for all monopolar modes. For low values of $B$ ranging from $0.5$ to $2.0$ T the rule is always obeyed for all the excitations and it would be very interesting to test it experimentally. For certain limited values of $B\ge2.5$ T the jump rule collapses as shown for example in Fig. \ref{fig4} and Fig. \ref{fig5}. However the magnetic field values where this happens can vary depending on the particular mode. The reason why the Raman intensity jump rule collapses at specific values of the magnetic field is related to changes in the properties of the quantum dot electronic ground and excited states that are manifested in small values of the Raman matrix elements in association with cancellation effects in Eq. (\ref{eq1}). Doubtlessly this point deserves further theoretical and experimental analysis. The Raman intensity jump rule is not universal with respect to the external magnetic field at higher field values but it can be a useful tool for identifying the spin or charge nature of the electronic excitations in quantum dots.

Part of this work was performed at the Institute for Microstructural Sciences (IMS), National Research Council, Ottawa. A.D. acknowledges the hospitality and support of IMS.

\end{document}